\documentstyle[amssymb,preprint,prb,aps]{revtex}

\begin{document}
\draft
\title{Theory for effects of pressure on heavy-fermion alloys}
\author{Sun Zhang}
\address{Department of Physics, Nanjing University, Nanjing 210093, China}
\maketitle

\begin{abstract}
The effects of pressure on heavy-fermion alloys are studied in the framework
of Yoshimori-Kasai model under the coherent potential approximation. A
unified picture is presented for both the electron-type heavy-fermion
systems and the hole-type heavy-fermion systems. The density of states of $f$
electrons is calculated over the whole range of the doping concentration
under the applied pressure. The Kondo temperature, the specific-heat
coefficient, and the electrical resistivity are obtained, in agreement with
the experiments qualitatively. The contrasting pressure-dependent effects
for two types of heavy-fermion alloys are discussed to reveal the coherence
in the system under pressure.
\end{abstract}


\section{INTRODUCTION}

During the past two decades, considerable attention has been focused on
heavy-fermion (HF) systems.\cite{1,2,3,4,4.5} Generally speaking, HF systems
are a class of intermetallic compounds which contain a periodic array of
magnetic Kondo ions, involving rare earth (4$f$) or actinide (5$f$)
elements. At higher temperatures than the Kondo temperature $T_K$, the
localized magnetic moments behave essentially as independent impurities and
each $f$ electron becomes the scattering center in the Kondo effect. In this
temperature region, the characters of the $f$ electrons are similar to those
in dilute alloys. At low temperatures, the coherent heavy fermion (or Fermi
liquid) behaviors are observed, the $f$ electrons form a coherent Kondo
lattice. The specific-heat coefficient $\gamma $ is of order $10^2\sim 10^3$
larger than that of normal metals, and the density of states near Fermi
level, deduced from $\gamma $, is enhanced enormously. It is widely accepted
that the large specific-heat coefficient is caused by Kondo effect at each
Kondo ion site.\cite{1}

From the ensuing experimental and theoretical works, it is clear that HF\
materials such as CeAl$_3$,\cite{5,6,7,8} CeCu$_6$,\cite{8,9,10,11,12,13} UBe%
$_{13}$,\cite{13,14,15,15.5} and CeCu$_2$Si$_2$,\cite{16,17} display many of
the characteristics of metallic Kondo lattice (KL), where a lattice of
localized magnetic moments coexists with a conduction band.\cite{18} Instead
of a Kondo resonance (single-peak) structure in the impurity case, the
density of states of $f$ electrons ($f$-DOS) has a pseudogap (two-peak)
structure near Fermi level in the KL case, due to the periodic coherence
between Kondo ions in the lattice. In order to get information on the
development of coherence in HF systems, many experiments have been performed
to study the alloying effects in doped HF systems, such as Ce$_x$La$_{1-x}$Cu%
$_2$Si$_2$,\cite{19,20,21} Ce$_x$La$_{1-x}$Al$_3$,\cite{22} Ce$_x$La$_{1-x}$%
Be$_{13}$,\cite{23} Ce$_x$La$_{1-x}$Cu$_6$,\cite{24,25,26,27} and so on. It
is shown that, with the increasing of the concentration of Kondo ions
(Ce-like ions), an HF alloy system undergoes a crossover from Kondo impurity
state to KL state. That is, doping Kondo ions presents a consecutive
approach to the coherent Kondo state efficiently. Meanwhile, the electrical
resistivity $\rho $ follows the quadratic law on temperature $T$ in KL case,
which is one of the characteristic features of coherence, corresponding to
the Fermi liquid behavior. Additionally, a shift of the peak in the
specific-heat coefficient $\gamma $ is observed to finite temperature with
increasing of the concentration $x$,\cite{21,22} while the Kondo temperature 
$T_K$ is independent of $x$ from experiments.\cite{20,24,27} Theoretically,
a dispersion term of $f$ electron band is introduced to the Anderson lattice
model (ALM) by Yoshimori and Kasai to get a metallic KL system in the case
of half-filling.\cite{28} Under the coherent potential approximation (CPA),
they calculate the electrical resistivity of HF alloys, and some excellent
results are obtained, in agreement with experiments very well. Furthermore,
Li and Qiu\cite{50} extent the method of Yoshimori and Kasai, and tried to
present a reasonable picture for heavy-fermion alloys by the application of
the slave-boson mean-field approximation. They calculated the density of
states, the thermoelectric power, the specific-heat coefficient, and the
residual resistivity. The results themselves look meaningful and the
obtained Kondo temperature $T_K$ is indeed independent of the alloy
concentration. While the slave-boson parameter has not been calculated out
over the whole range of the concentration $0\leqslant x\leqslant 1$, their
results seem artificial and more careful study is necessary to be performed
even for the alloying effects of HF systems.

On the other hand, it is also possible to study systematically the
development of the HF behaviors through the application of pressure. From
the pressure-dependent measurements on CeAl$_3$,\cite{5,6,7,8} CeCu$_6$,\cite
{8,9,10,11,12,13} UBe$_{13}$,\cite{13,14,15} CeInCu$_2$,\cite{29,30,31,32,33}
YbCu$_2$Si$_2$,\cite{34,35} YbCuAl,\cite{35,36} etc., it is clear that, for
Ce-based and U-based HF systems, the specific-heat coefficient is depressed
and the temperature $T_{\max }$, at which the electronic resistivity gets
the maximum, shifts to higher temperature under pressure. Furthermore,
pressure tends to expand the temperature region of the quadratic law and to
enhance the coherence of the system. While for Yb-based HF systems, just the
opposite effects appear under pressure. More generally, pressure
qualitatively acts as a {\it mirror} between Ce-based, U-based compounds and
Yb-based compounds.\cite{34,35} From experiments, pressure destabilizes the
larger ion. In the case of $f$ electron systems, pressure will stabilize the 
$f^{n-1}$ configuration relative to $f^n$.\cite{15,37a,37b} For Ce-based
compounds, the nonmagnetic $4f^0$ configuration would be favored rather than 
$4f^1$ under pressure.\cite{13} While for Yb-based compounds, the magnetic $%
4f^{13}$ configuration would be favored under pressure rather than $4f^{14}$.%
\cite{13,37c} Moreover, pressure tends to induce a crossover from a
localized $f$-electron states to an itinerant or coherent one for Ce-based
compounds,\cite{30,33} and the opposing effect is also expected for Yb-based
compounds.\cite{34,13} Although these important results stated above have
been obtained from experiments for a long time, to the best of our
knowledge, few works have been performed for the pressure effects on the HF
systems theoretically. Among various possible approaches to study the
pressure effects, the Kondo collapse (KC) is considered as a reasonable
mechanism. It can be used to explain such fact that the hybridization $V$ is
very exponentially sensitive to pressure. But the KC theory also indicates
that the volume decrease results in a large increase in the Kondo
temperature.\cite{18} It is true for the Ce-based\cite{30,31,32,33} and
U-based\cite{14,15} compounds, but not true for the Yb-based compounds where
the volume decrease results in a decrease in the Kondo temperature.\cite{34}
So such a mechanism, the KC, is still controversial, and it seems not so
efficient to think that the mirror effect between Ce-based, U-based
compounds and Yb-based compounds can be explained within the framework of
the KC. The essential of the problem is how pressure influences the
development of coherence, in which the system displays typical HF features.
Taking into account the valence fluctuation of $f$ ions between the singlet
and magnetic multiplet configurations, the pressure-dependent behaviors may
arise from the cell-volume difference between $f$ configurations. Pressure
favors the $f$ configuration with a smaller volume.\cite{37,38} On this
basis, we would like to develop a theory of pressure on HF alloys involving
effects of doping and effects of pressure simultaneously. Following Li and
Qiu,\cite{50} the alloying effects of HF systems are studied within the
Yoshimori-Kasai (YK) model by using the slave-boson mean-field approximation
(SBMFA), but the main results are calcultated again over the whole range of
the concentration $0\leqslant x\leqslant 1$ carefully in order to promote
the credit of the method. The aim of the paper is to present a unified
picture of pressure effects for both Ce-based, U-based and Yb-based systems.

The rest of the paper is prepared as follows. In Sec. II, we formulate the
disorder scattering within the mean-field approximation of YK model and
introduce the volume variable to describe the pressure influence via the $f$
valence fluctuation, originating from the hybridization between $f$
electrons and conduction electrons ($c$ electrons). Then the spectral
function of single-particle Green's function (GF) is obtained using CPA
method and a set of self-consistent equations is also addressed. Based on
these, the $f$-DOS is performed numerically in Sec. III, while the
specific-heat coefficient and the electrical resistivity are discussed in
detail in Sec. IV, where we attempt to explain and compare the contrasting
effects of pressure on Ce-based, U-based and Yb-based systems in a unified
theory. Finally, our results are summarized and discussed in Sec. V.

\section{CPA DISORDER FORMALISM AND PRESSURE MODEL FOR HF ALLOYS}

In the alloy systems such as Ce$_x$La$_{1-x}$Cu$_2$Si$_2$, and Ce$_x$La$%
_{1-x}$Al$_3$, there exists two kinds of rare earth atoms $A$ and $B$, where 
$A$ (Ce-like) is a magnetic atom with $f$ electrons and $B$ (La-like) a
nonmagnetic atom without $f$ electrons. The substitution of an $A$ atom by a 
$B$ atom introduces the disorder into the system and creates missing $f$
electrons, referred to as Kondo holes. Following Li and Qiu,\cite{50} the
random variable at the lattice point $l$ is defined by 
\begin{equation}
\xi _l=\left\{ 
\begin{array}{l}
1\ \qquad \qquad \text{for }l\in A, \\ 
0\ \qquad \qquad \text{for }l\in B,
\end{array}
\right.  \label{z1}
\end{equation}
and $\overline{\xi _l}=x$, the normalized concentration of $A$ atoms. The YK
model is the Anderson lattice model, with a small dispersion on $f$ electron
band,\cite{28} and the disorder Hamiltonian can be written as 
\begin{eqnarray}
H &=&\sum_{{\bf k}m}[\varepsilon _{{\bf k}}c_{{\bf k}m}^{\dagger }c_{{\bf k}%
m}+(\alpha \varepsilon _{{\bf k}}-E_0)f_{{\bf k}m}^{\dagger }f_{{\bf k}%
m}]+\sum_{lm}(1-\xi _l)(E_L+E_0)f_{lm}^{\dagger }f_{lm}  \nonumber \\
&&+V\sum_{lm}\xi _l(f_{lm}^{\dagger }c_{lm}+c_{lm}^{\dagger }f_{lm})+\frac 12%
U\sum_{l,m\neq m^{\prime }}\xi _lf_{lm}^{\dagger }f_{lm}f_{lm^{\prime
}}^{\dagger }f_{lm^{\prime }},  \label{z2}
\end{eqnarray}
where ($-E_0$) and $E_L$ are the energy levels of $f$ electrons on the $A$
sites and $B$ sites, respectively. $\varepsilon _{{\bf k}}$ is the energy of 
$c$ electrons from Fermi level, which is assumed to be zero. $U$ gives the
on-site Coulomb repulsion between $f$ electrons and $V$ the $c$-$f$ mixing
parameter. The dispersion term $\alpha \varepsilon _{{\bf k}}$ is introduced
into the YK model to get the metallic states even in the case of
half-filling and the parameter $\alpha $ is proportional to $V^2$. Other
notions in Eq. (\ref{z2}) are standard.

In the strong correlation limit $U\rightarrow \infty $, double occupation on 
$A$ sites is forbidden and the Coleman's slave-boson (SB) operator $b_l$ is
introduced in the $c$-$f$ mixing term.\cite{39} Then the YK Hamiltonian (\ref
{z2}) in SB formalism reads 
\begin{eqnarray}
H &=&\sum_{{\bf k}m}[\varepsilon _{{\bf k}}c_{{\bf k}m}^{\dagger }c_{{\bf k}%
m}+(\alpha \varepsilon _{{\bf k}}-E_0)f_{{\bf k}m}^{\dagger }f_{{\bf k}%
m}]+\sum_{lm}(1-\xi _l)(E_L+E_0)f_{lm}^{\dagger }f_{lm}  \nonumber \\
&&+V\sum_{lm}\xi _l(b_lf_{lm}^{\dagger }c_{lm}+c_{lm}^{\dagger
}f_{lm}b_l^{\dagger })+\sum_l\xi _l\lambda _l\left( \sum_mf_{lm}^{\dagger
}f_{lm}+b_l^{\dagger }b_l-1\right) ,  \label{z3}
\end{eqnarray}
where a constraint 
\begin{equation}
\sum_mf_{lm}^{\dagger }f_{lm}+b_l^{\dagger }b_l=1\qquad \text{for }l\in A,
\label{z4}
\end{equation}
has been added with the Lagrange multiplier $\lambda _l$. Such a constraint
prevents the double occupancy of $f$ level on $A$ sites due to the infinite $%
U$.

In order to consider the effects of pressure, let us introduce the total
volume operator. In pure KL systems ($x=1$), such as CeCu$_6$, UBe$_{13}$,
YbCuAl etc., there is a lattice of rare earth or actinide ions which can
exist in two valence states:\cite{39,40} One of them is typically a singlet, 
$f^n(j=0)$ with zero $j$; the other a $2j+1(=N)$-fold-degenerate state, $%
f^{n+1}(j,+m)$ or $f^{n-1}(j,-m)$ with spin $j$. The weak hybridization
between $c\ $electrons and the local $f$ electrons causes the valence to
fluctuate by the following changes in the $f$ shell occupation: 
\begin{eqnarray}
f^{n+1}(j,+m) &\rightleftharpoons &f^n(j=0)+e^{-}(j,m)\hspace{3cm}\text{for
Ce and U,}  \label{x5} \\
f^{n-1}(j,-m) &\rightleftharpoons &f^n(j=0)+h^{+}(j,m)\hspace{3cm}\text{for
Yb.}  \label{x6}
\end{eqnarray}
According to the SB technique of Coleman\cite{39} 
\begin{equation}
\mid f^n;j=0\rangle _l\equiv b_l^{\dagger }\mid 0\rangle _l,  \label{z7}
\end{equation}
\begin{equation}
\mid f^{n\pm 1};j,\pm m\rangle _l\equiv f_{lm}^{\dagger }\mid 0\rangle _l.
\label{z8}
\end{equation}
Then at each site $l$, the valence fluctuation can be represented by a
resonance between a zero-energy boson and a spin-$j$ fermion in the subspace
where $Q=n_b+n_f=1$. The fermion is an electron $e^{-}$ for Ce and U, while
a hole $h^{+}$ for Yb, respectively. In this paper, the numbers of channel $%
N(=2j+1)=2$ would be taken for simplicity, and two values ($\pm \frac 12$)
are considered for $m$, written as $\sigma $ from now on.

Taking into account the cell-volume difference $\Delta \Omega =\Omega
_1-\Omega _0$ between two $f$ configurations, we can write down the total
volume operator as\cite{37} 
\begin{equation}
\Omega _t=\sum_l\Omega _l=\sum_l[b_l^{\dagger }b_l\Omega _0+(1-b_l^{\dagger
}b_l)\Omega _1],  \label{z9}
\end{equation}
where $\Omega _0$ and $\Omega _1$ are the cell volume for the singlet $f^n$ (%
$b_l^{\dagger }b_l=1$) and the multiplet states $f^{n\pm 1}$($b_l^{\dagger
}b_l=0$), respectively. Then, $\Delta \Omega $ is either positive for the
cells with the electron-type ($e$-type) $f$ ions (Ce and U) or negative for
the cells with the hole-type ($h$-type) $f$ ions (Yb). The more electrons
occupy the $f$ shell, the larger the ionic radius is.

In the case of alloy ($0\leq x\leq 1$), we can express the total volume
operator in terms of the random variable $\xi _l$ as\cite{38} 
\begin{equation}
\Omega _t=\sum_l\{(1-\xi _l)\Omega _L+\xi _l[b_l^{\dagger }b_l\Omega
_0+(1-b_l^{\dagger }b_l)\Omega _1]\},  \label{z10}
\end{equation}
where $\Omega _L$ is the cell volume of a Kondo hole site (with La-like
ions), $\Omega _0$ and $\Omega _1$ are the cell volumes of an $A$ atom in
singlet $f^n$ ($b_l^{\dagger }b_l=1$) and multiplet $f^{n\pm 1}(b_l^{\dagger
}b_l=0)$ states, respectively.

In the SBMFA, the operator $b_l$ and constraint (\ref{z4}) are replaced by
their mean-field values with the ansatz $r=\left\langle b_l\right\rangle $
and $\lambda =\lambda _l$ for all $A$ sites. Then the mean-field Hamiltonian
is 
\begin{eqnarray}
H_{MF} &=&\sum_{{\bf k}\sigma }[\varepsilon _{{\bf k}}c_{{\bf k}\sigma
}^{\dagger }c_{{\bf k}\sigma }+(\alpha \varepsilon _{{\bf k}}+E_f)f_{{\bf k}%
\sigma }^{\dagger }f_{{\bf k}\sigma }]+\sum_{l\sigma }(1-\xi _l)\varepsilon
_Lf_{l\sigma }^{\dagger }f_{l\sigma }  \nonumber \\
&&+rV\sum_{l\sigma }\xi _l(f_{l\sigma }^{\dagger }c_{l\sigma }+c_{l\sigma
}^{\dagger }f_{l\sigma })+x\lambda N_s(r^2-1),  \label{z11}
\end{eqnarray}
where $E_f=\lambda -E_0$ and $\varepsilon _L=E_L-E_f$ are the renormalized $%
f $ level of the magnetic ($A$) atoms and the Kondo holes ($B$ atoms),
respectively. Here we have used the relation $x=N_s^{-1}\sum_l\xi _l$, and $%
N_s$ is the total number of sites in the system.

From Eq. (\ref{z10}), the averaged cell-volume is 
\begin{equation}
\overline{\Omega }_l=(1-x)\Omega _L+x[\Omega _0+(1-r^2)\Delta \Omega ].
\label{x12}
\end{equation}
Since pressure always decreases the averaged cell-volume $\overline{\Omega }%
_l$, for the $e$-type HF systems (such as CeCu$_6$ and UBe$_{13}$) where $%
\Delta \Omega >0$, pressure will lead to the increasing of $r^2$. While for
the $h$-type HF systems (such as YbCuAl), an opposite effect appears since $%
\Delta \Omega <0$.

Because a Kondo hole doping will lead to a very strong scattering, $%
E_L\rightarrow \infty $, so that $\varepsilon _L\rightarrow \infty $. We
should solve the disorder slave-boson mean-field Hamiltonian (\ref{z11}) for
arbitrary concentrations by means of a nonperturbative approach, the CPA.%
\cite{41,42,43} Here, we would like to give the full steps to get the
analytic solution of the coherent potential, instead of the procedure by Li
and Qiu,\cite{50} where such a solution is introduced directly.

To perform the CPA, we should introduce a translational invariant but
frequency-dependent coherent potential of the effective medium to replace
the disorder scattering potential in Hamiltonian (\ref{z11}). The coherent
potential for a $c$-$f$ mixing model such as YK model could be assumed as a $%
2\times 2$ matrix\cite{50,38,44a} 
\begin{equation}
S(\omega ,x)=\left( 
\begin{array}{ll}
S_{cc} & \quad S_{cf} \\ 
S_{fc} & \quad S_{ff}
\end{array}
\right) ,  \label{z12}
\end{equation}
and the average site Green's function (GF) of the effective medium is
obtained 
\begin{equation}
F(\omega )=\frac 1{N_s}\sum_{{\bf k}}\overline{G}(\omega ,\text{ }{\bf k}%
)=\left( 
\begin{array}{ll}
F_{cc}(\omega ) & F_{cf}(\omega ) \\ 
F_{fc}(\omega ) & F_{ff}(\omega )
\end{array}
\right) .  \label{z16}
\end{equation}

Then the effective medium Hamiltonian can be written in the matrix form 
\begin{equation}
\overline{H}=\sum_{{\bf k}\sigma }( 
\begin{array}{ll}
c_{{\bf k}\sigma }^{\dagger } & f_{{\bf k}\sigma }^{\dagger }
\end{array}
)\left( 
\begin{array}{ll}
\varepsilon _{{\bf k}}+S_{cc} & \text{ \quad \ \ \thinspace }S_{cf} \\ 
\text{ \ \ \thinspace }S_{fc} & \quad (\alpha \varepsilon _{{\bf k}%
}+E_f)+S_{ff}
\end{array}
\right) \left( 
\begin{array}{l}
c_{{\bf k}\sigma } \\ 
f_{{\bf k}\sigma }
\end{array}
\right) +x\lambda N_s(r^2-1).  \label{z13}
\end{equation}

From the difference between the disorder Hamiltonian (\ref{z11}) and the
effective medium Hamiltonian (\ref{z13}), 
\begin{equation}
H_{MF}-\overline{H}=\sum_lV_l,  \label{z19}
\end{equation}
the scattering potentials for atoms $A$ and $B$ are reached, 
\begin{equation}
V_A=\left( 
\begin{array}{ll}
\text{ \ }-S_{cc} & \text{ \ \ }rV-S_{cf} \\ 
rV-S_{fc} & \text{ \ \ \ }-S_{ff}
\end{array}
\right) ,\qquad V_B=\left( 
\begin{array}{ll}
-S_{cc} & \text{ \ \ \thinspace \thinspace \thinspace }-S_{cf} \\ 
-S_{fc} & \text{ \ \ }\varepsilon _L-S_{ff}
\end{array}
\right) .  \label{z20}
\end{equation}

According to Yonezawa,\cite{43} the self-consistent condition in single-site
CPA is 
\begin{equation}
xt_A+(1-x)t_B=0,  \label{z17}
\end{equation}
where $t_A$ and $t_B$ are the scattering $t$ matrices for $A$ and $B$ atoms,
respectively, 
\begin{equation}
t_{A(B)}=V_{A(B)}[1-F(\omega )V_{A(B)}]^{-1}.  \label{z18}
\end{equation}

From Eqs. (\ref{z20}), (\ref{z17}), and (\ref{z18}), and taking $\varepsilon
_L\rightarrow \infty $ to ensure no $f$ electron occupation on Kondo holes,
we can find an analytic solution of the coherent potential 
\begin{equation}
S(\omega ,x)=\left( 
\begin{array}{ll}
0 & \quad rV \\ 
rV & \quad S_{ff}
\end{array}
\right) ,  \label{z21}
\end{equation}
where $S_{cc}=0$, $S_{cf}=S_{fc}=rV$ and only $S_{ff}$ is to be determined.
At the same time, the scattering $t$ matrices can be simplified as 
\begin{equation}
t_A=\frac 1{1+S_{ff}F_{ff}}\left( 
\begin{array}{ll}
0 & \text{ \ \quad }0 \\ 
0 & \text{ \ }-S_{ff}
\end{array}
\right) ,\qquad t_B=\frac 1{F_{ff}}\left( 
\begin{array}{ll}
0 & \text{ \ \ \thinspace \thinspace \thinspace }0 \\ 
0 & \text{ \ \ }-1
\end{array}
\right) ,  \label{x21}
\end{equation}
and the self-consistent CPA equation (\ref{z17}) can be written as 
\begin{equation}
S_{ff}F_{ff}=x-1.  \label{z22}
\end{equation}
Now, the analytic solution of $S(\omega ,x)$ and the self-consistent CPA
equation are obtained after the detail derivation. These expressions are the
same as those in the paper by Li and Qiu,\cite{50} but the approach to them
are not given there. Then, the average site GFs are expressed as 
\begin{equation}
F_{cc}(\omega )=\frac 1{N_s}\sum_{{\bf k}}\frac{\omega -\alpha \varepsilon _{%
{\bf k}}-E_f-S_{ff}}{(\omega -\varepsilon _{{\bf k}})(\omega -\alpha
\varepsilon _{{\bf k}}-E_f-S_{ff})-(rV)^2},  \label{z23}
\end{equation}
\begin{equation}
F_{cf}(\omega )=F_{fc}(\omega )=\frac 1{N_s}\sum_{{\bf k}}\frac{rV}{(\omega
-\varepsilon _{{\bf k}})(\omega -\alpha \varepsilon _{{\bf k}%
}-E_f-S_{ff})-(rV)^2},  \label{z24}
\end{equation}
\begin{equation}
F_{ff}(\omega )=\frac 1{N_s}\sum_{{\bf k}}\frac{\omega -\varepsilon _{{\bf k}%
}}{(\omega -\varepsilon _{{\bf k}})(\omega -\alpha \varepsilon _{{\bf k}%
}-E_f-S_{ff})-(rV)^2},  \label{z25}
\end{equation}
which are also very different from those in the paper by Li and Qiu.\cite{50}

The parameters of SB, $r$ and $\lambda $, can be determined by the extreme
values of the grand canonical free enthalpy's variations (or equivalently,
by Hellmann-Feynman theorem). The grand canonical free enthalpy of the HF
alloy system under pressure $p$ is 
\begin{equation}
K=-\beta ^{-1}\ln Z_{MF},  \label{z26}
\end{equation}
where 
\begin{equation}
Z_{MF}=\text{Tr}\{\exp [-\beta (\overline{H}+p\Omega _t)]\}\equiv \text{Tr}%
[\exp (-\beta H_{eff})].  \label{z27}
\end{equation}
It is easy to write down the effective Hamiltonian of the SBMFA 
\begin{eqnarray}
H_{eff} &=&\sum_{{\bf k}\sigma }[\varepsilon _{{\bf k}}c_{{\bf k}\sigma
}^{\dagger }c_{{\bf k}\sigma }+(\alpha \varepsilon _{{\bf k}}+E_f+S_{ff})f_{%
{\bf k}\sigma }^{\dagger }f_{{\bf k}\sigma }+rV(f_{{\bf k}\sigma }^{\dagger
}c_{{\bf k}\sigma }+c_{{\bf k}\sigma }^{\dagger }f_{{\bf k}\sigma })] 
\nonumber \\
&&+(1-x)N_sp\Omega _L+xN_s\{\lambda (r^2-1)+p[\Omega _0+(1-r^2)\Delta \Omega
]\}.  \label{z28}
\end{eqnarray}
From the variation with respect to $\lambda $, 
\begin{equation}
0=\frac{\delta K}{\delta \lambda }=\left\langle \frac{\partial H_{eff}}{%
\partial \lambda }\right\rangle _T=xN_s(r^2-1)+\sum_{{\bf k}\sigma }\langle
f_{{\bf k}{\Bbb \sigma }}^{\dagger }f_{{\bf k}{\Bbb \sigma }}\rangle _T,
\label{z29}
\end{equation}
we get the equation including the parameter $r$, 
\begin{equation}
x(1-r^2)=\frac 1{N_s}\sum_{{\bf k}\sigma }\langle f_{{\bf k}{\Bbb \sigma }%
}^{\dagger }f_{{\bf k}{\Bbb \sigma }}\rangle _T=-\frac 2\pi \int_{-\infty
}^\infty d\omega f(\omega )%
\mathop{\rm Im}%
F_{ff}(\omega +i0^{+}).  \label{z30}
\end{equation}
And application of the same procedure to $r$, 
\begin{eqnarray}
0 &=&\frac{\delta K}{\delta r}=\left\langle \frac{\partial H_{eff}}{\partial
r}\right\rangle _T  \nonumber \\
&=&V\sum_{{\bf k}\sigma }(\langle c_{{\bf k}\sigma }^{\dagger }f_{{\bf k}%
\sigma }\rangle _T+\langle f_{{\bf k}\sigma }^{\dagger }c_{{\bf k}\sigma
}\rangle _T)+2xN_sr(\lambda -p\Delta \Omega ),  \label{z31}
\end{eqnarray}
implies another equation 
\begin{eqnarray}
xr(\lambda -p\Delta \Omega ) &=&-\frac V{N_s}\sum_{{\bf k}\sigma }\langle f_{%
{\bf k}\sigma }^{\dagger }c_{{\bf k}\sigma }\rangle _T  \nonumber \\
&=&\frac{2V}\pi \int_{-\infty }^\infty d\omega f(\omega )%
\mathop{\rm Im}%
F_{fc}(\omega +i0^{+}).  \label{z32}
\end{eqnarray}

Eqs. (\ref{z22}), (\ref{z23}), (\ref{z24}), (\ref{z25}), (\ref{z30}), and (%
\ref{z32}) constitute a set of self-consistent equations. These equations
are not only fundamental to determine the coherent potential $S_{ff}(\omega
,x)$ and the SB parameter $r$ of the HF alloy systems, but also powerful for
calculating the electronic DOS of both $c$ electrons and $f$ electrons with
arbitrary alloy concentration under various applied pressures.

\section{DENSITY OF STATES AND THE KONDO TEMPERATURE}

The $f$-DOS per magnetic ($A$) site for each spin is defined by 
\begin{equation}
N_f(\omega ,\text{ }p\Delta \Omega \text{, }x)=-\frac 1{\pi x}\text{Im}%
F_{ff}(\omega +i0^{+}),  \label{z36}
\end{equation}
where $F_{ff}$ can be calculated self-consistently by numerical method. In
the calculations, the unperturbed DOS of conduction electrons, $N_0(\omega )$%
, is assumed as\cite{44} 
\begin{equation}
N_0(\omega )=\frac 2{\pi D}\sqrt{1-\left( \frac \omega D\right) ^2}\Theta
\left( D-\left| \omega \right| \right) ,  \label{z37}
\end{equation}
where $\Theta (x)$ is the step function and $D$ the half-width of the
unperturbed conduction band.

In the case of a half-filled conduction band, $r^2\ll 1$. From Eq. (\ref{z4}%
), $n_f\lesssim 1$ is obtained, corresponding to the Kondo limit associated
with a large $E_0$.\cite{40} Therefore, $\lambda $ can be expected to be of
order $E_0$,\cite{45} and we can take $E_f=0$ and $\lambda =E_0$ as a
reasonable approximation. The phenomenological parameter $\alpha $ in the
dispersion term\cite{46,47,48} can be written as\cite{51} 
\begin{equation}
\alpha =\eta ^2%
{rV \overwithdelims() D}%
^2,  \label{z38}
\end{equation}
where the parameter $\eta $ is greater than 1 in order to get a metallic KL
model without real gap.\cite{50}

The numerical results of $f$-DOS are performed at various concentrations
under applied pressure, shown in Fig. 1. It is clear that no real energy gap
appears in the DOS of electrons for arbitrary concentration, and the
metallic behavior is obtained.

On the one hand, from Fig. 1, with the increasing of the concentration $x$,
the $f$-DOS transforms from a Kondo impurity resonant state with single-peak
structure into a Kondo coherent state with two-peak pseudogap structure. In
the dilute region ($x\rightarrow 0$) only a single peak appears, indicating
the local impurity $f$ states, the system behaves as a collection of
independent Kondo singlet on each $A$ site and every $f$ electron becomes
the scattering center in the Kondo effect. When the concentration of $f$
ions increases, the delocalization of the $f$ electrons enhances due to the
growing coherent scattering. After the concentration reaches a critical
value, $x_c\lesssim 0.7$, for the parameters chosen for calculation here,
the curve of $f$-DOS near $\omega =0$ transforms into the concave shape from
a convex one, showing the appearance of the coherent (two-peak) pseudogap
state. In the high concentration region ($x\rightarrow 1$), the $f$
electrons form a coherent lattice, a global Kondo singlet occurs, and the
collection of independent Kondo singlets in dilute limit is replaced by a
whole coherent Kondo lattice state here.

On the other hand, if pressure is applied, two different effects are reached
for the $e$-type HF alloys ($\Delta \Omega >0$) and the $h$-type HF alloys ($%
\Delta \Omega <0$). For the $e$-type alloys, pressure increases the Kondo
interaction and correlation between $f$ ions ($A$ sites),\cite{13,14,15} and
tends to enhance the itinerance of $f$ electrons and the coherence of the
system.\cite{30,33} The width of the $f$-DOS near $\omega =0$ is broadened
and the height is lowered under pressure. While for the $h$-type alloys, the
opposing effect occurs. Pressure decreases the Kondo interaction and
correlation between $f$ ions, and tends to reduce the itinerance of $f$
electrons and the coherence of the system.\cite{13,34} The width of the $f$%
-DOS near $\omega =0$ is depressed and the height is lifted under pressure.

The effective mixing parameter $(rV)^2$ is also calculated out, shown in
Fig. 2, over the whole range of alloy concentration $0\leq x\leq 1$ under
various pressures. On the one hand, with the increasing concentration of $x$%
, $(rV)^2$ increases linearly and connects two well-known mean-field
results, the Kondo impurity ($x\rightarrow 0$) and the Kondo lattice ($%
x\rightarrow 1$) naturally.\cite{50} On the other hand, as shown in Fig. 2,
the effect of pressure appears sensitively in the change of the effective
mixing parameter. It is in agreement with experiments\cite{13,34}
qulitatively that pressure tends to promote increase in hybridization and
the mixing parameter but decrease in localization for the $e$-type HF alloys
($\Delta \Omega >0$), while conversely, for the $h$-type HF alloys ($\Delta
\Omega <0$).

Furthermore, we would like to give some discussion on the pressure effects
of the Kondo temperature. The Kondo temperature $T_K$ is an energy scale,
which is used to characterize the contribution to resistivity $\rho $ with
the temperature relation $\rho (T)=a-b\ln (T)$ due to the inelastic
scattering of conduction electrons from partially compensated local moments.
With the decreasing of the temperature, a crossover occurs from the
incoherent Kondo impurity state ($T>T_K$) to the coherent KL state ($T<T_K$%
). Theoretically, $T_K$ can be determined in the limit $r\rightarrow 0$.
From the self-consistent equations, it is directly obtained 
\begin{equation}
\frac{2V^2}{(\lambda -p\Delta \Omega )\pi }\int_{-\infty }^\infty d\omega 
\frac 1{\exp \left( \frac \omega {k_BT_K}\right) +1}\frac 1{N_s}\sum_{{\bf k}%
}\text{Im}\frac 1{(\omega -\varepsilon _{{\bf k}}+i0^{+})(\omega -E_f+i0^{+})%
}=1,  \label{x39}
\end{equation}
and the expression for $T_K$ reads 
\begin{equation}
T_K=1.13\frac D{k_B}\exp \left( -\frac{E_0-p\Delta \Omega }{2N_0(0)V^2}%
\right) .  \label{x40}
\end{equation}
This analytic expression for the Kondo temperature is given here explicitly
as one of the important results of pressure effects. We would like to point
out that such an expression does not appear in the paper by Li and Qiu,\cite
{50} where the case $p=0$ is considered and no effect of pressure is
obtained.

From Eqs. (\ref{x39}) and (\ref{x40}), it is clear that, the Kondo
temperature $T_K$ is independent of the concentration $x$, in agreement with
the experiments.\cite{20,24,27} On the other hand, as shown in Fig. 3,
pressure increases the Kondo temperature $T_K$ for the $e$-type HF systems ($%
\Delta \Omega >0$), which is in agreement with the experiments on UBe$_{13}$,%
\cite{14,15} CeInCu$_2$,\cite{30,31,32,33} and other $e$-type materials.
While for the $h$-type HF systems ($\Delta \Omega <0$), $T_K$ decreases with
the increment of pressure. Some experiments also indicate that the Kondo
temperature $T_K$ seems to decrease with pressure in the $h$-type compounds.%
\cite{34}

According to the discussion about the $f$-DOS, the effective mixing
parameter $(rV)^2$, and the Kondo temperature $T_K$, it is believed that
pressure does act as a {\it mirror} between the $e$-type HF systems and the $%
h$-type HF systems unambiguously, in agreement with the experiments. Based
on these results, we would calculate the specific-heat coefficient and the
electrical resistivity to discuss the behaviors of HF system under pressure
more intensively.

\section{COHERENCE EFFECTS UNDER PRESSURE}

In the low temperature region $T\ll T_K$, the main contribution of the
specific heat arises from the thermal activation of $f$ electrons near Fermi
level, and the specific-heat coefficient $\gamma $ of HF alloys can be
written in terms of $f$-DOS as\cite{50} 
\begin{equation}
\gamma (T,\text{ }p\Delta \Omega ,\text{ }x)=\frac 12k_B^2\beta
^3\int_{-\infty }^\infty d\omega \omega ^2N_f(\omega ,\text{ }p\Delta \Omega
,\text{ }x)\text{sech}^2\left( \frac{\beta \omega }2\right) ,  \label{x41}
\end{equation}
where $\beta =1/k_BT$. Then, from the $f$-DOS\ given above, the
specific-heat coefficient $\gamma $ can be obtained over the whole range of
concentration $x$ under various applied pressures (Fig. 4), the influence of
pressure on $T_K$ is also considered. It is shown that when $x<x_c$, the
case of Kondo impurity, $\gamma (T)$ continues to increase monotonically as $%
T$ decreases, indicating the incoherent state of the alloy system.\cite{21}
While in the KL case ($x>x_c$), a maximum is found at a finite temperature
well below $T_K$ and a peak appears in each $\gamma $-$T$ curve. With the
increasing of concentration $x$, the peak shifts to higher temperature
corresponding to the pseudogap structure. Generally, the maximum value of $%
\gamma (T)$, found at a finite temperature, is a characteristic feature of
KL, which signals the transition to the coherent state and can be assigned
experimentally to the effect of the periodicity of the system.\cite{1,21,22}

On the other hand, as to the effect of pressure on the specific-heat
coefficient $\gamma $, two opposing results are shown in Fig. 4. For the $e$%
-type HF alloys ($\Delta \Omega >0$), pressure increases the correlation
between $f$ ions and promotes the itinerance of $f$ electrons. The global
Kondo singlet tends to be stabilized and the $f$ multiplet states
suppressed. Then, near Fermi level, the $f$-DOS is lowered and the thermal
activation decreases, leading to the decrement of $\gamma $ under pressure.
These results are in agreement with the measurements on the $e$-type HF
systems CeAl$_3$,\cite{7,8} CeCu$_6$,\cite{12} UBe$_{13}$,\cite{15.5,8} and
CeCu$_2$Si$_2$.\cite{16} For the $h$-type HF alloys ($\Delta \Omega <0$),
pressure promotes the localization of $f$ electrons, and tends to
destabilize the global singlet state. The $f$-DOS near Fermi level is lifted
and the thermal activation increases, leading to the increment of $\gamma $
under pressure. These results are opposite to those of the $e$-type HF
systems, as observed in YbCuAl,\cite{36} etc.

From experiments, accompanying the decrease (increase) of $\gamma $ for the $%
e$-type ($h$-type) HF systems is a rapid suppression (enhancement) of the $%
T^2$ coefficient $A$ of the electrical resistivity $\rho $, an increase
(decrease) of the temperature interval over which $\rho =\rho _0+AT^2$ and
an increase (decrease) of the temperature $T_{\max }$ at which $\rho $ has
its maximum.\cite{34,8,9,13} Study on the electrical resistivity $\rho $
presents an effective approach to reveal the effect of coherence in HF
systems. According to Kubo formula, the CPA expression of the electrical
resistivity can be written as\cite{42} 
\begin{equation}
\sigma (T,\text{ }p,\text{ }x)=\frac{2e^2v_F^2}{3\pi \hbar ^2\Omega }%
\int_{-\infty }^\infty d\omega \left( -\frac{\partial f}{\partial \omega }%
\right) \sum_{{\bf k}}[\text{Im}\overline{G}_{cc}({\bf k},\text{ }p,\text{ }%
\omega +i0^{+})]^2,  \label{z42}
\end{equation}
where $v_F$ is the Fermi velocity, $\Omega $ the volume of the system and 
\begin{equation}
\overline{G}_{cc}({\bf k},\text{ }p,\text{ }\omega )=\frac{\omega -\alpha
\varepsilon _{{\bf k}}-E_f-S_{ff}}{(\omega -\varepsilon _{{\bf k}})(\omega
-\alpha \varepsilon _{{\bf k}}-E_f-S_{ff})-(rV)^2},  \label{z43}
\end{equation}
is the matrix element of the effective medium GF for conduction electrons.

At $T=0$, the residual resistivity $\rho _0(T=0,p,x)=1/\sigma (T=0,p,x)$ are
calculated as shown in Fig. 5. It is found that $\rho _0$ has a maximum
within $0\leqslant x\leqslant 1$, and approximately follows the Nordheim law 
$\rho _0\propto x(1-x)$.\cite{20,23,24,27,28} On the other hand, for the $e$%
-type HF systems, $\rho _0$ decreases with increasing pressure,
corresponding to the decrease in effective mass due to the enhancement of
itinerance and coherence of the system under pressure,\cite{30} which is in
agreement with the observation in UBe$_{13}$,\cite{13,14,15} CeCu$_6$,\cite
{9,13} CeCu$_2$Si$_2$,\cite{35} CeAl$_3$,\cite{5,6} CeInCu$_2$,\cite
{30,31,32,33} etc. While for the $h$-type HF systems, $\rho _0$ increases
with increasing pressure, corresponding to the increase in effective mass
due to the enhancement of localization and incoherent state of the system
under pressure,\cite{13} as observed in YbAgCu$_4$,\cite{34} YbCu$_2$Si$_2$,%
\cite{34} etc. In the KL case ($x=1$), the $\rho $-$T$ curves are given in
Fig. 6. The resistivity follows the quadratic law $\rho =\rho _0+AT^2$
(here, $\rho _0=0$ when $x=1$) at low temperature and has a maximum at $%
T_{\max }$. The coefficient $A$ (Fig. 7) and the temperature $T_{\max }$
(Fig. 8) are strongly affected by the application of pressure. For the $e$%
-type HF systems, pressure increases the temperature $T_{\max }$ but
decreases the coefficient $A$ and expands the temperature region in which
the quadratic law appears. Because the larger presence of the quadratic term
in the temperature dependence of the resistivity accompanies the stronger
coherence, these results indicate that pressure studies on the $e$-type HF
systems provide a means of tuning the onset of Kondo coherence into the
experimental temperature range without the introduction of disorder which
accompanies doping.\cite{14} For the $h$-type HF systems, pressure decreases 
$T_{\max }$ but increases the coefficient $A$ and shrinks the temperature
range for the quadratic law. Again, pressure acts as a mirror and leads to
the contrasting effects on resistivity between the $e$-type HF systems as UBe%
$_{13}$,\cite{13,14,15} CeCu$_6$,\cite{9,13} CeCu$_2$Si$_2$,\cite{35} CeAl$%
_3 $,\cite{5,6} and CeInCu$_2$,\cite{30,31,32,33} and the $h$-type HF
systems as YbAgCu$_4$,\cite{34} YbCuAl,\cite{35} and YbCu$_2$Si$_2$.\cite
{34,35}

\section{CONCLUSIONS}

In this paper, the pressure effects on HF alloys are studied by the
application of CPA in the framework of Yoshimori-Kasai (YK) model. Following
Li and Qiu, the alloying effects of HF systems are studied by using the
SBMFA. The density of states of $f$ electrons ($f$-DOS), the Kondo
temperature $T_K$, the specific-heat coefficient $\gamma $, and the
electrical resistivity $\rho $ are obtained in our CPA formalism for both
the $e$-type HF alloys and the $h$-type HF alloys. It is found that with the
increasing of the $f$ ion concentration, as pointed out by Li and Qiu,\cite
{50} the system transforms from a Kondo impurity with single-peak structure
into a coherent Kondo lattice (KL) with two-peak pseudogap structure.
Accompanying this transformation, a peak appears in the specific-heat
coefficient $\gamma $, and it shifts to higher temperature by increasing
doping. On the other hand, pressure tends to enhance the itinerance and
coherence of the system in the $e$-type HF alloys, while, to support the
localization and the incoherent state in the $h$-type HF alloys. The
application of pressure increases the Kondo temperature and suppresses the
specific-heat coefficient for the $e$-type HF systems. Accompanying these
effects, pressure increases the temperature $T_{\max }$ and expands the
temperature region for the quadratic law. Conversely, pressure decreases the
Kondo temperature, enhances the specific-heat coefficient and shrinks the
temperature interval for the quadratic law in the $h$-type case. Our
theoretical results on the HF systems can be looked as a unified
interpretation on the opposite pressure-dependent effects, observed in UBe$%
_{13}$, CeCu$_6$, CeCu$_2$Si$_2$, CeAl$_3$, and CeInCu$_2$ (the $e$-type HF
systems) and YbAgCu$_4$, YbCuAl, and YbCu$_2$Si$_2$ (the $h$-type HF
systems). Although, it is widely accepted that pressure acts qualitatively
as a mirror between Ce-based, U-based compounds and Yb-based compounds,\cite
{34,35} further experiments are required to determine to what intensity and
to what extent the $h$-type HF systems are mirror images of their $e$-type
counterparts.

\section{ACKNOWLEDGMENT}

The author is very grateful to Prof. Z. Z. Li and Dr. Y. Qiu for the
stimulating viewpoints in their original work, which benefit the author
greatly.

\begin{figure}[tbp]
\caption{Pressure effect on the $f$-DOS\ of HF alloys, for $x=0$, $0.4$, $%
0.8 $, and $1.0$. The parameters for the numerical calculation are $%
V^2=0.2D^2$, $E_0=1.2D$, and $\eta =1.05$.}
\label{1}
\end{figure}

\begin{figure}[tbp]
\caption{Pressure effect on the effective mixing parameter $(rV)^2$ for the $%
e$-type ($\Delta \Omega >0$) and the $h$-type ($\Delta \Omega <0$) HF alloys
over the whole range of $x$ under various pressures.}
\label{2}
\end{figure}

\begin{figure}[tbp]
\caption{Pressure effect on the Kondo temperature $T_K$ for the $e$-type ($%
\Delta \Omega >0$) and the $h$-type ($\Delta \Omega <0$) HF alloys.}
\label{3}
\end{figure}

\begin{figure}[tbp]
\caption{Pressure effect on the specific-heat coefficient $\gamma $ of HF
alloys for $x=0$, $0.4$, $0.8$, and $1.0$.}
\label{4}
\end{figure}

\begin{figure}[tbp]
\caption{Pressure effect on the residual resistivity $\rho _0$ over the
whole range of the concentration $x$, where $\rho _a$ is the maximum value
of the curve $p=0$.}
\label{5}
\end{figure}

\begin{figure}[tbp]
\caption{Pressure effect on the electrical resistivity $\rho $ of Kondo
lattice ($x=1$), where $\rho _u=3\pi \hbar ^2D^2\Omega /2e^2v_F^2$ and $T_0$
is the $T_{\max }$ at $p=0$.}
\label{6}
\end{figure}

\begin{figure}[tbp]
\caption{Pressure effect on the coefficient $A$ of the quadratic law of the
resistivity in unit of $A_0$ for the Kondo model ($x=1$), where $A_0$ is the
value of the $A$ at $p=0$.}
\label{7}
\end{figure}

\begin{figure}[tbp]
\caption{Pressure effect on the temperature $T_{\max }$ at which $\rho $ has
its maximum for the Kondo model ($x=1$), where $T_0$ is the $T_{\max }$ at $%
p=0$.}
\label{8}
\end{figure}

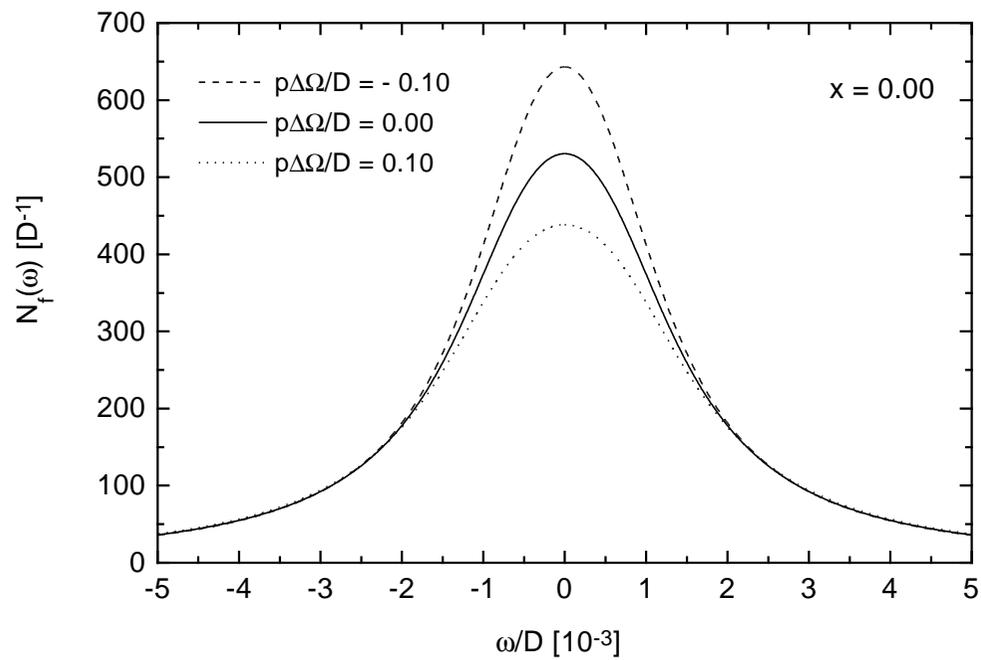
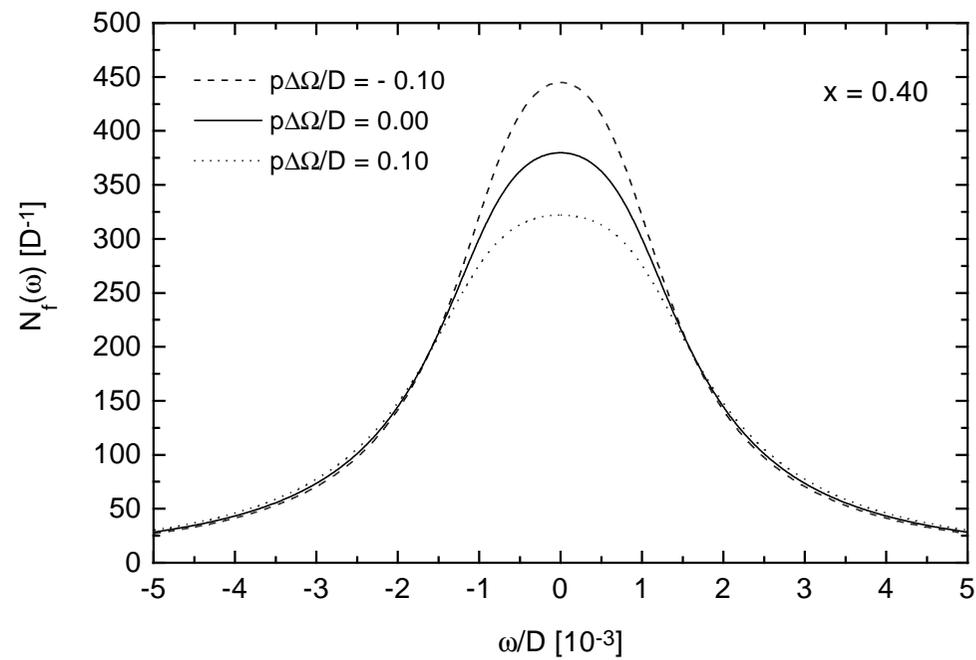
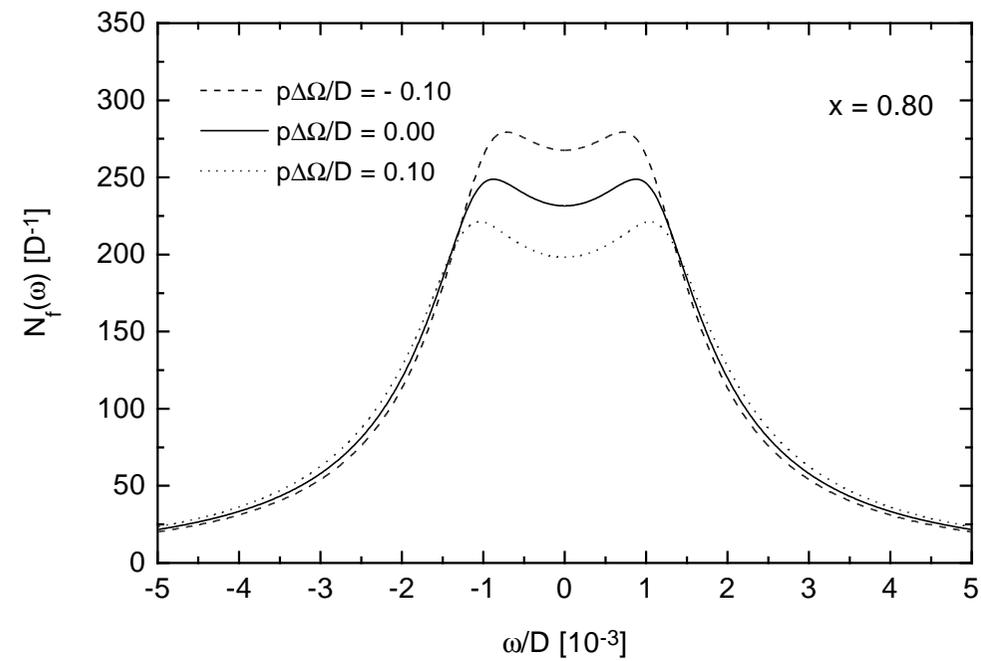
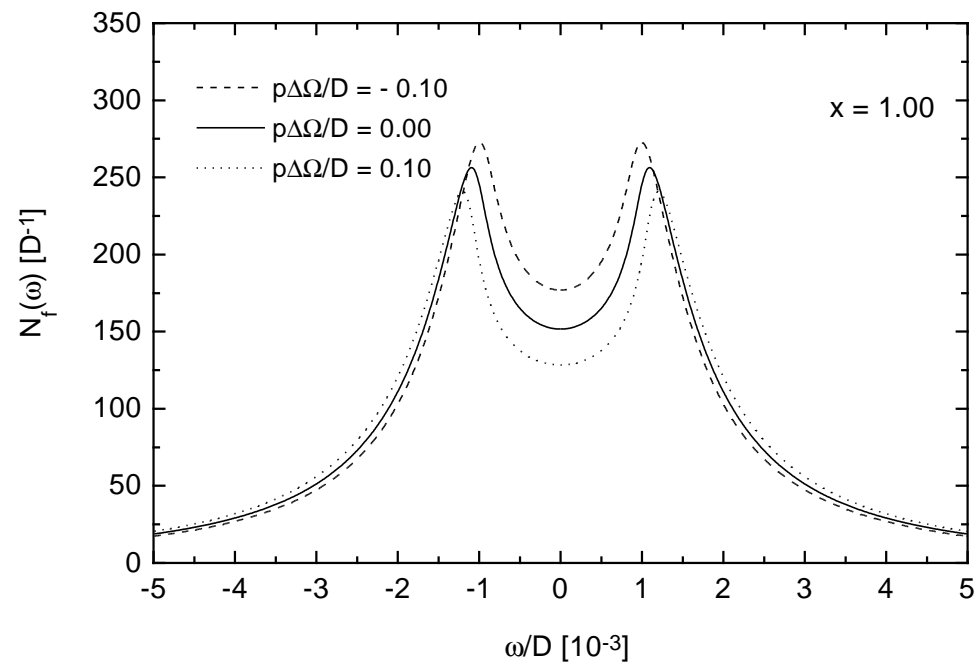
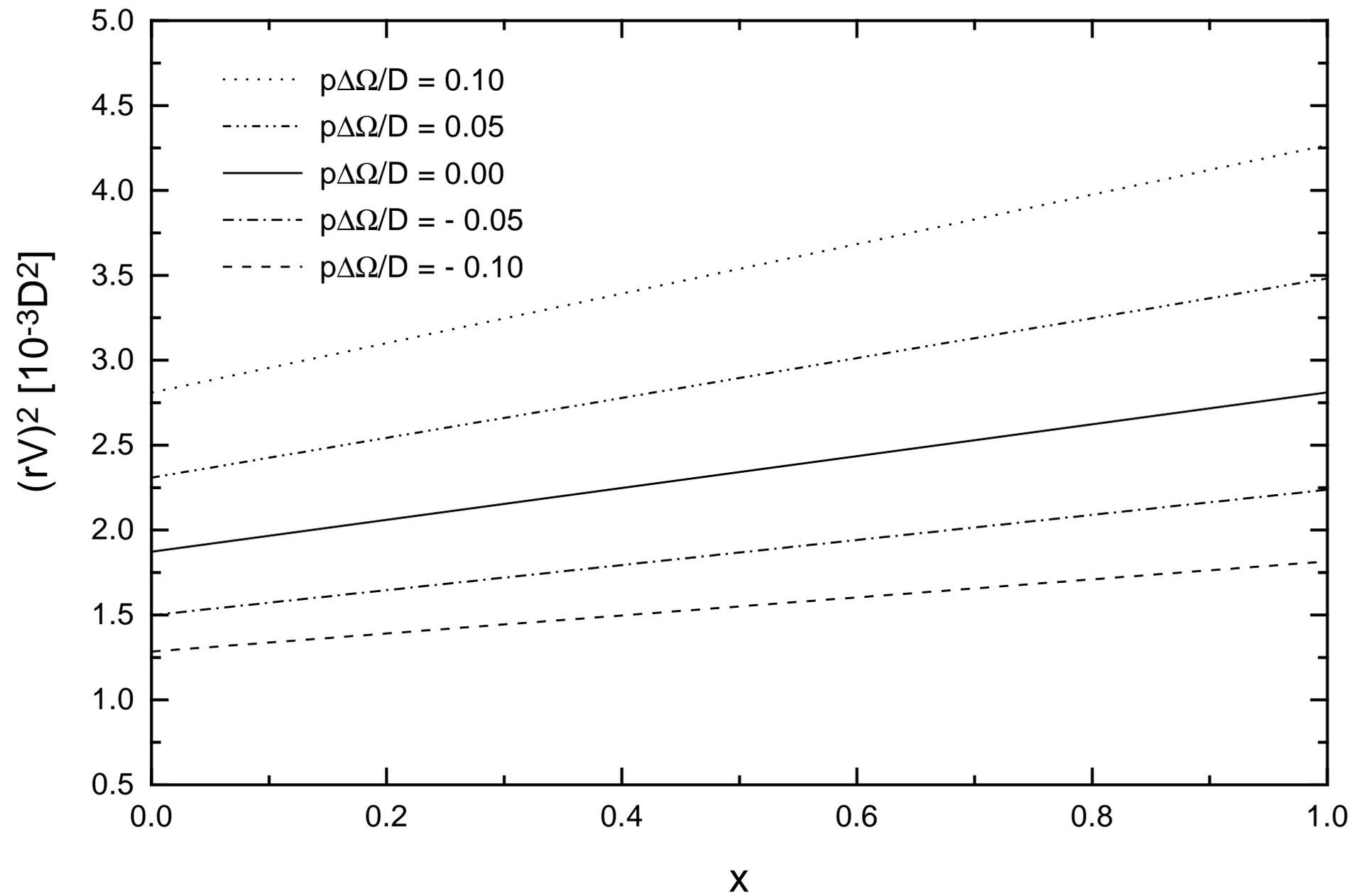
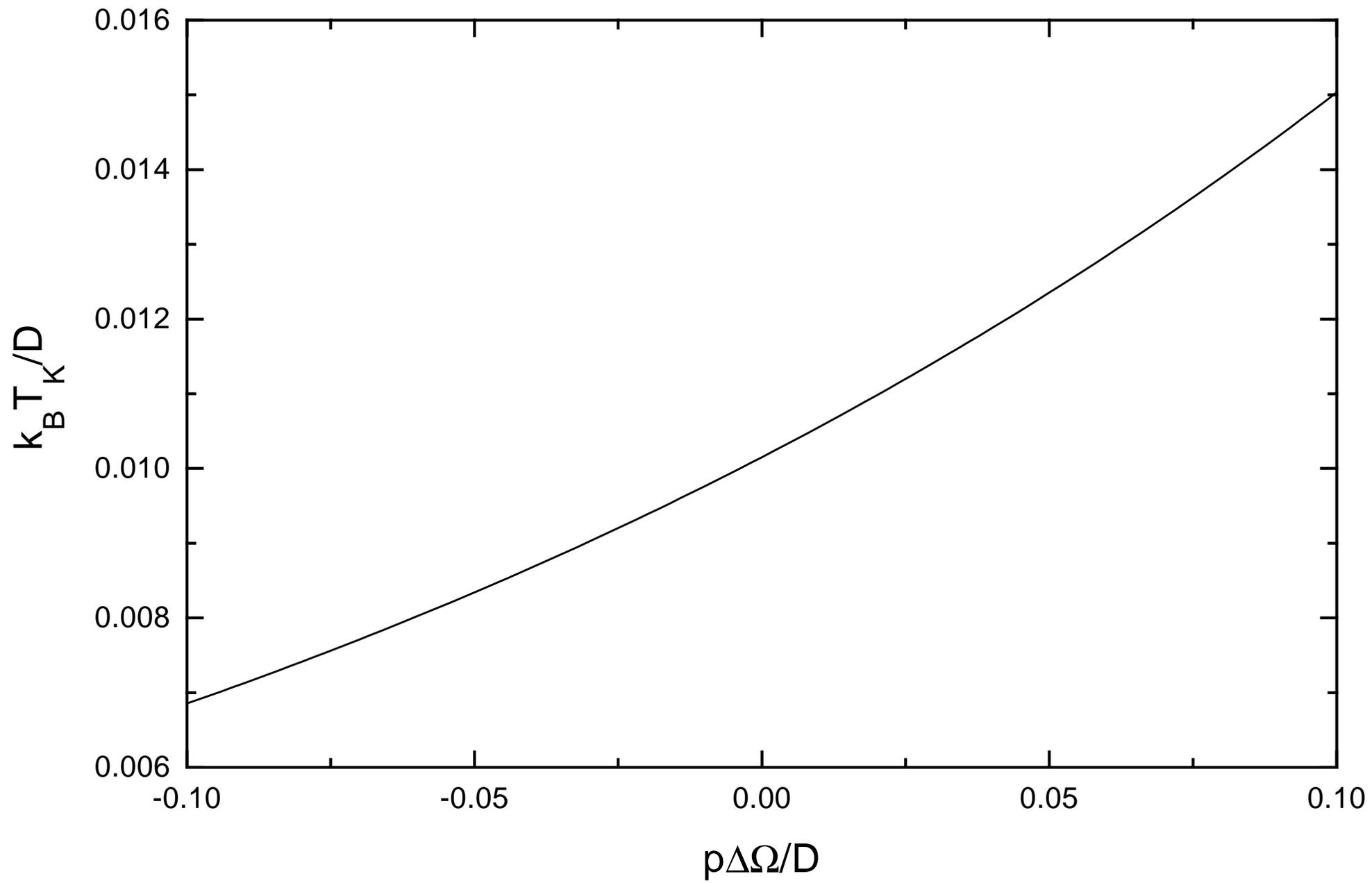
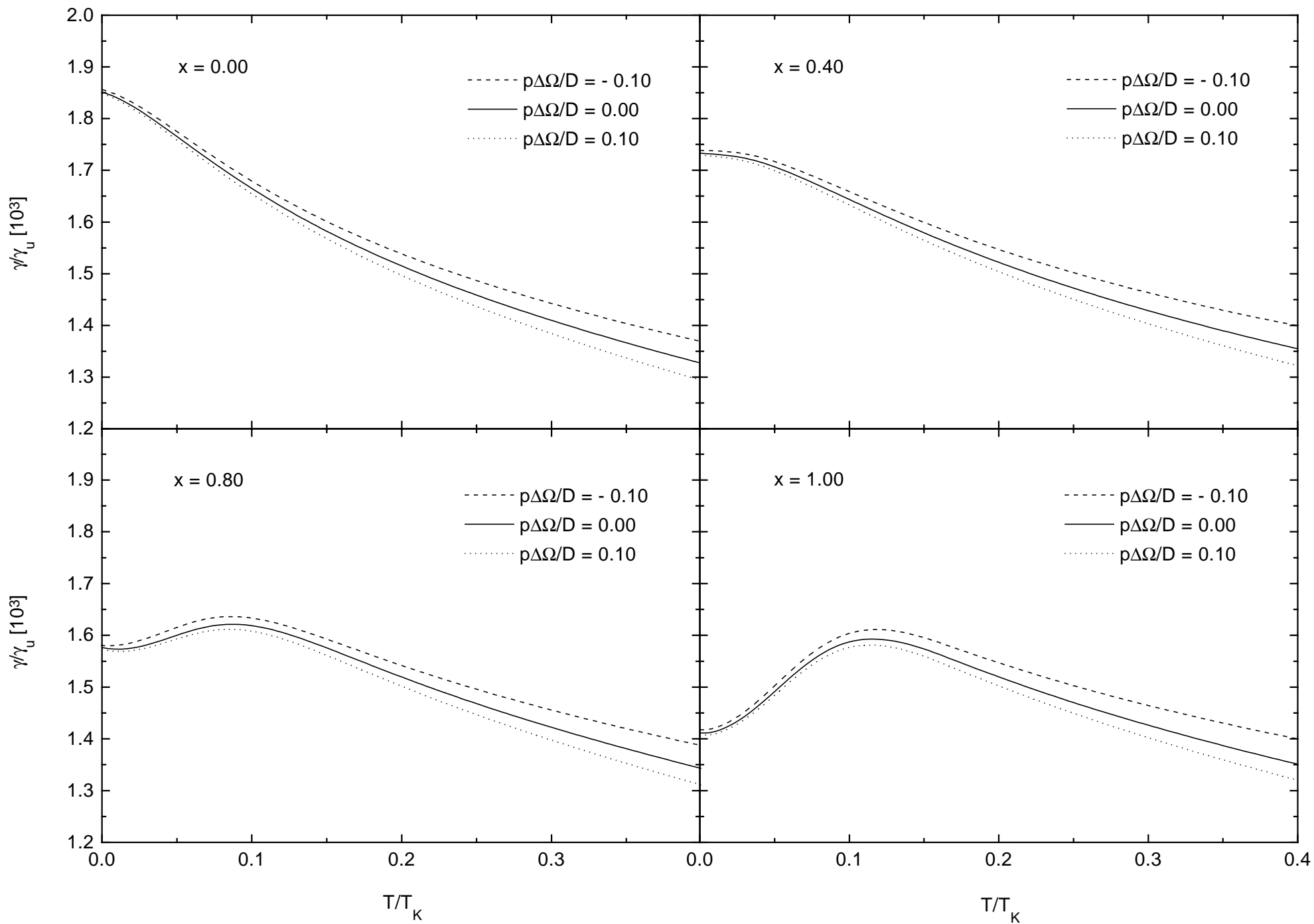
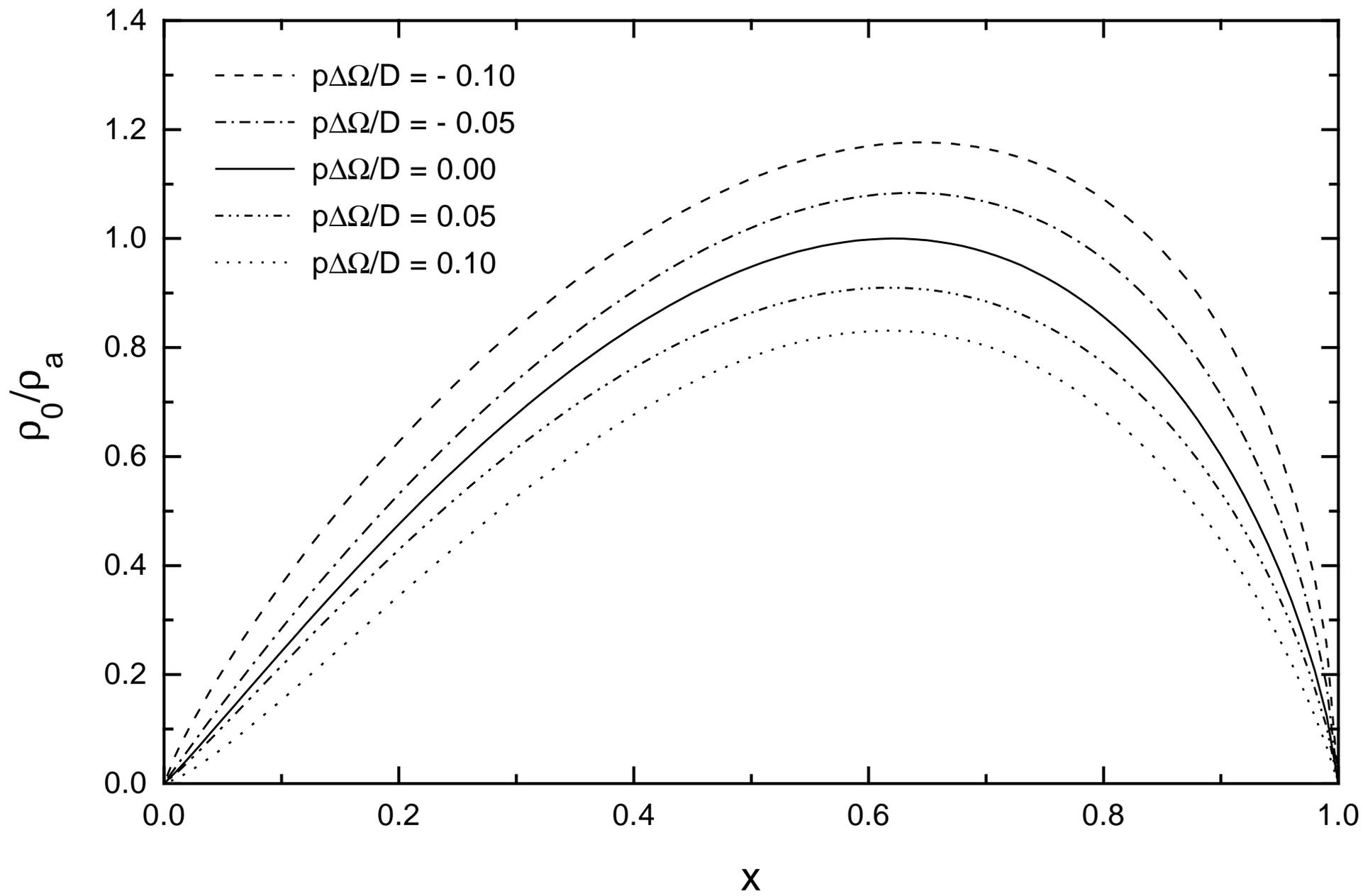
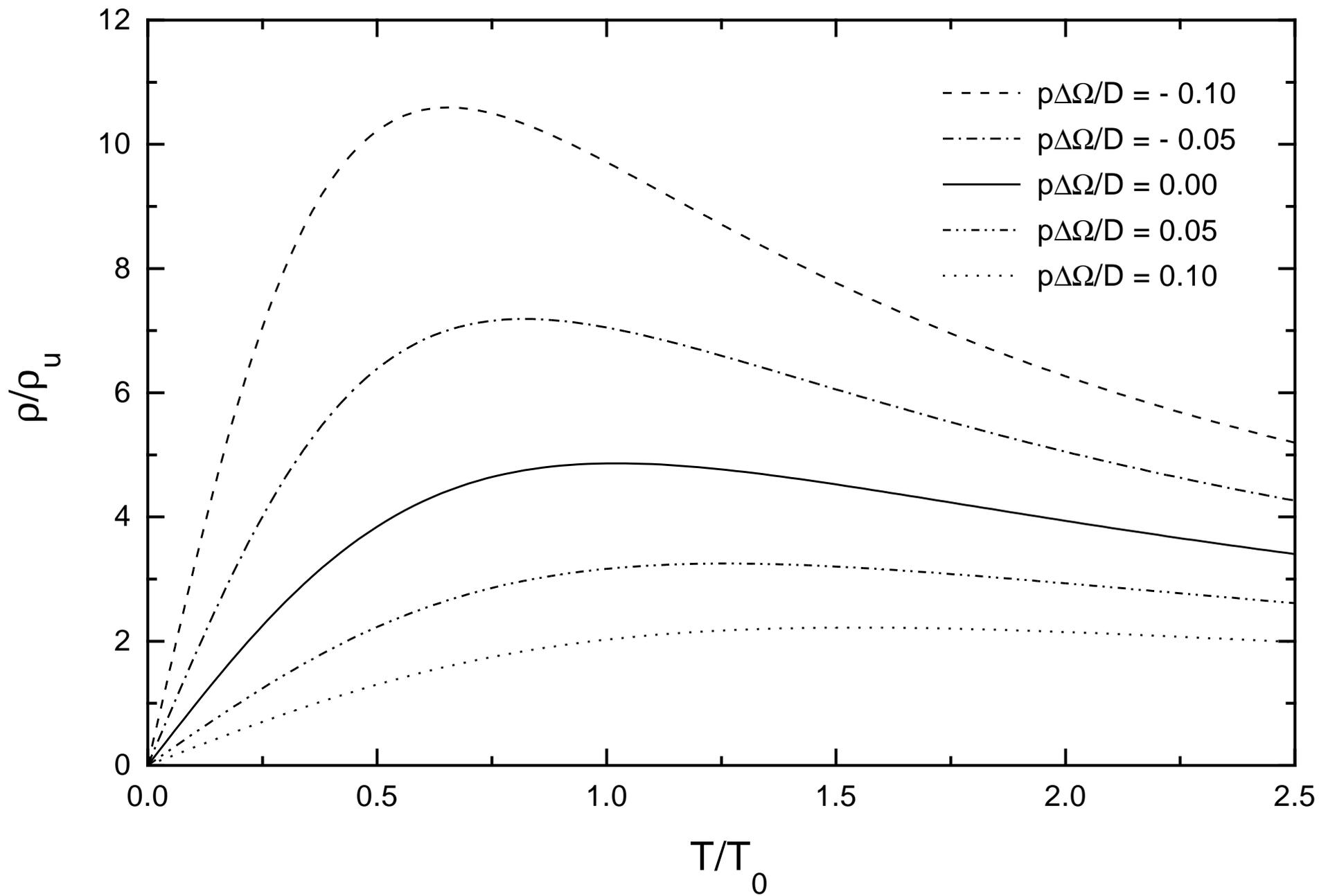
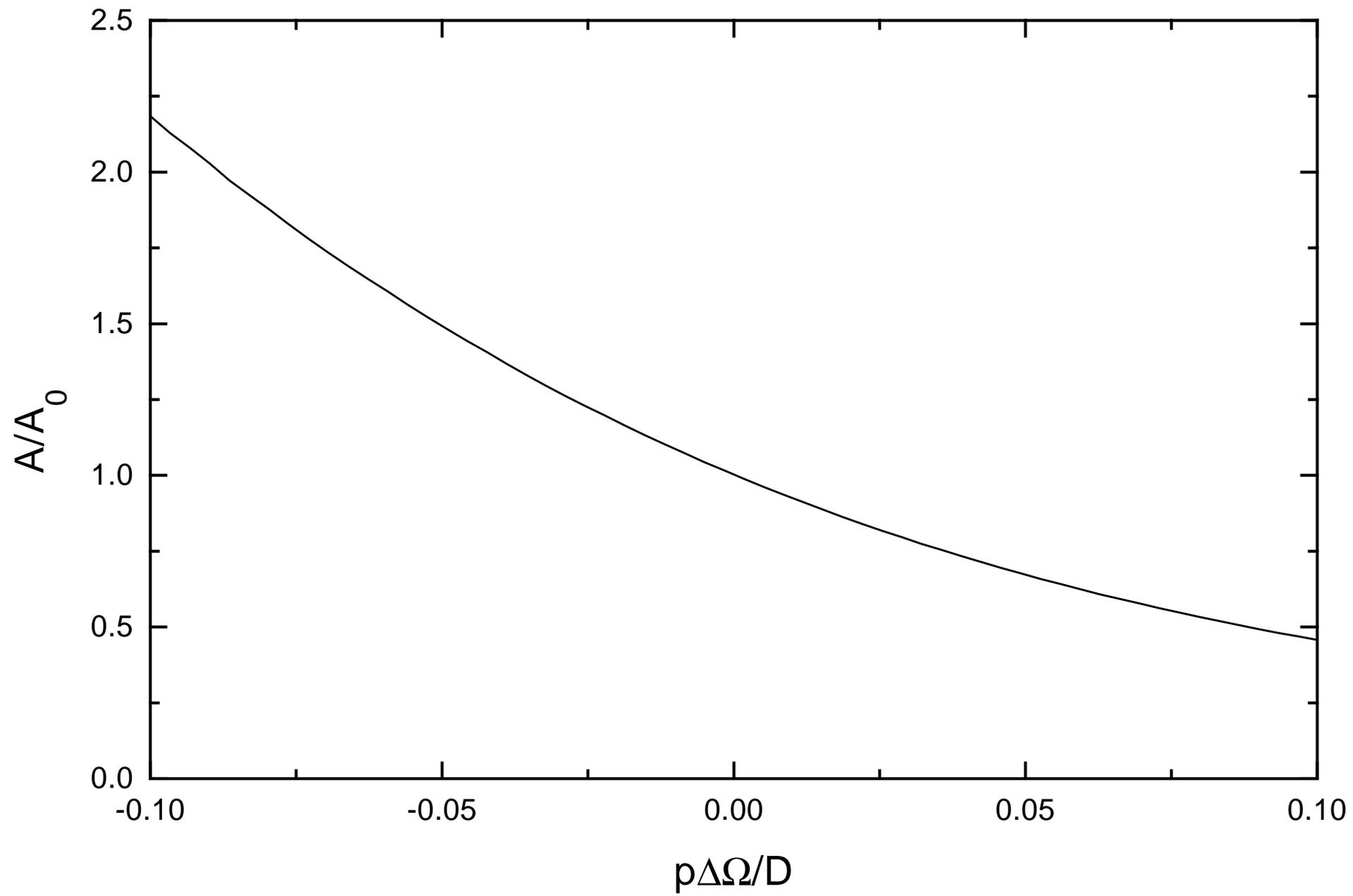
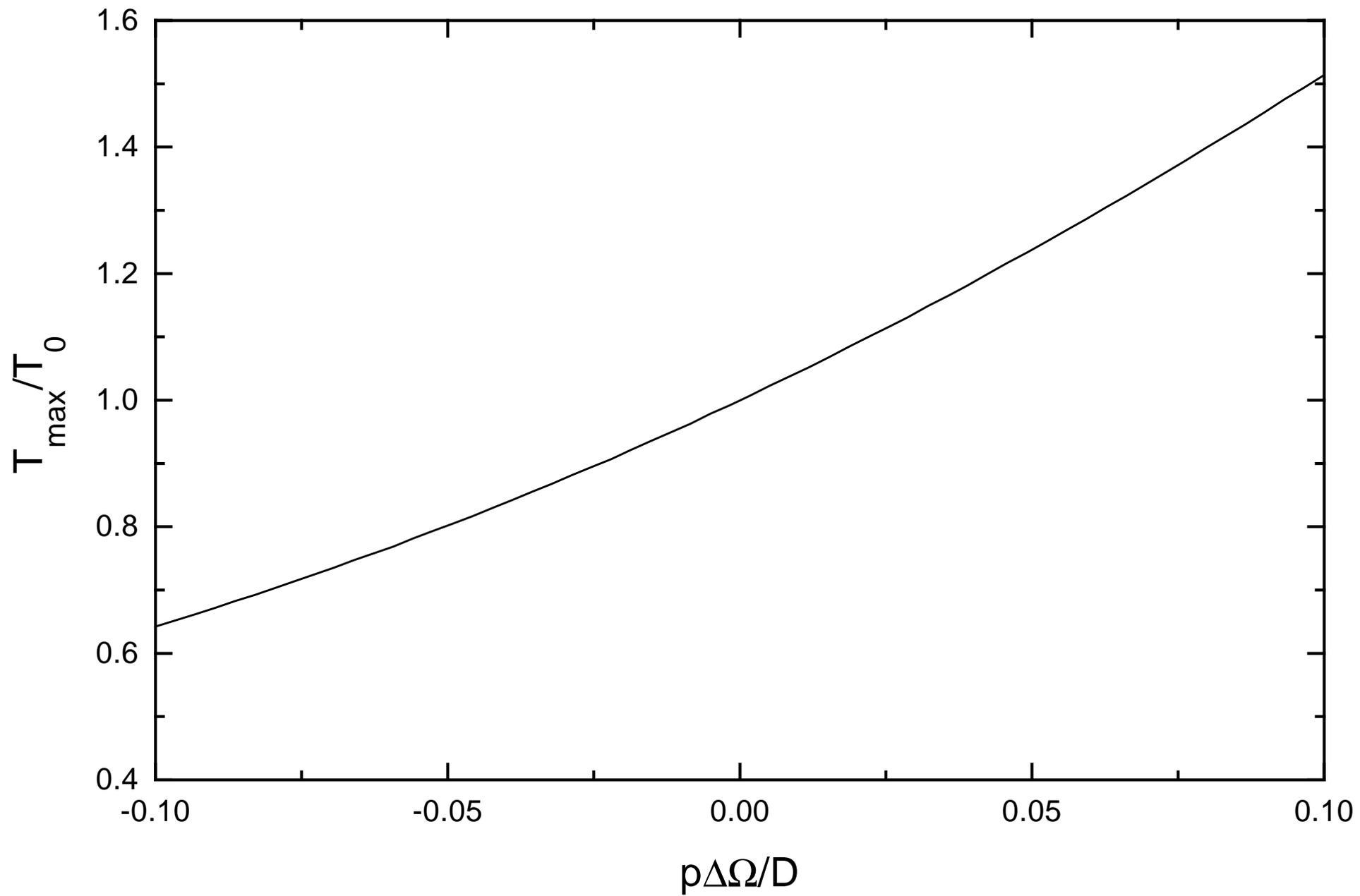
\end{document}